# A Secure RFID Deactivation/Activation Mechanism for Supporting Customer Service and Consumer Shopping

Jun-Chao Lu[1], Yu-Yi Chen[2], Zhen-Jie Qiu[1], Jinn-Ke Jan[1]

[1] Computer Science and Engineering, National Chung-Hsing University, Taiwan

{phd9416,s9756028, jkjan }@cs.nchu.edu.tw

[2] Management Information Systems, National Chung-Hsing University, Taiwan

chenyuyi@nchu.edu.tw

*Abstract*—RFID has been regarded as a time and money-saving solution for a wide variety of applications, such as manufacturing, supply chain management, and inventory control, etc. However, there are some security problems on RFID in the product managements. The most concerned issues are the tracking and the location privacy. Numerous scholars tried to solve these problems, but their proposals do not include the after-sales service. In this paper, we propose a purchase and after-sales service RFID scheme for shopping mall. The location privacy, confidentiality, data integrity, and some security protection are hold in this propose mechanism.

## 1. Introduction

RFID has been regarded as a time and money-saving solution for a wide variety of applications, such as manufacturing, supply chain management, and inventory control, etc. However, there are some security problems on RFID in the product managements. The most concerned issues are the tracking and the location privacy.

According to John Ayoade's definition [1], location threat is placing covert readers at specific locations creates two types of privacy threats. First, individuals carrying unique tags can be monitored and their location revealed if the monitoring agency knows the tags associated with those individuals. Second, a tagged object's location – regardless of who (or what) is carrying it – is susceptible to unauthorized disclosure. According above definition, it's possible that the carried tag is activated for someone's location can be targeted according to the unique tag ID in clothing or shoes. Moreover, activating tag could also raise some customer's privacy problems. For example, consumer's medications and other contents of a pocket could be readable and the amount of money in a wallet might be easily calculable by a reader. One woman's dress size could be publicly realized by any nearby readers. The privacy issues of RFID tags have raised recent attention in the popular press. More and more such kinds of negative news stories make the clothing retailer Benetton to withdraw plans for embedding RFID tags in its products of apparel. [2]

The most straightforward approach to protect consumer privacy is that the shops would "kill" the tags of purchased goods. The killed tags are no more functional and never be re-activated. The most well-known example is the "kill command" specified by EPCglobal[5], which permits the deactivation of tags after purchasing. However, the kill command is too forceful and negative. The major problem after it has been killed is that the owners would never be able to take any advantages that might be provided by an RFID tag. It is an "all or nothing" privacy mechanism [14]. Supposed a tag is deactivated, it can never be used for after-sale purposes, no matter how useful and interesting they would be for the consumer, such as intelligent home appliances or emerging applications. For example, the deactivated tag can not be used for a microwave oven to read cooking directions from food packages.

Besides the kill function to disable the tag permanently, blocking tags' signal is noticed to be another fitting solution for the owner privacy protection. Ari Juels proposed a "blocker tag" mechanism [10] to prevent malicious attackers from locating the target tag. The blocker tag simulates all conceivable tag identities and feigns that all possible tags exist there, thus preventing the reader from identifying the tags. Another well-known Faraday cage approach, assumed an RFID tag could be shielded from surveillance by using Faraday cage, a container made of foil or metal mesh that is impassable by radio signals.

However, blocking the tags' signal will not always be a proper solution to avoid the tag being tracked as it is attached to a large scale object. Applying cryptographic mechanisms is mainstream research in the RFID identification presently. The most researches could be called the "mutual authentication" model, only the authenticated reader can get the corresponding response for identifying the tag. A proposed research

[3] categorized the mutual authentication model to Hash-based approach [12, 16, 17], probabilistic encryption approach [11], pseudonym approach [6, 15, 18], and Error correction codes (ECC) with secret parameters approach [3]. Furthermore, RFID "deactivation/activation" model is another kind of solution for against the illegal tracking. As the tag's privacy flag [9] is setting off, the tag keeps silent or just responds unquestioningly to all requests for non-tracking. The tag will be setting on under the owner's permission as necessary for the tag should be identifiable again. Of course, the flipping of privacy flag should be only controlled by some kinds of secure scheme.

Considering the RFID "deactivation/activation" model is a simple and efficient mechanism for against the illegal tracking, we will propose how to control the privacy flag is flipped under the owner's permission. Moreover, the support of after-sales service is also considered in our design. It is important since the shops offer the after-sales service have been proved to be not only vital parts of consumers' satisfaction, but also an essential fundamental of high revenues and profit [7].

## 2. The secure g consumer shopping and customer service scheme

In the total solutions of intelligent store [13, 19, 20], the store issues RFID membership cards, and the goods will also be embedded RFID tags for improving the stores' management. The intelligent shopping cart will be equipped the RFID reader for the consumer to inquire about goods' information by scan the attached RFID tags. The shopping scenarios can be referred to the figure 1: (a) The consumer uses his membership card to log on the shopping cart's reader. Then, the intelligent shopping cart can provide the location of goods listed in the consumer's shopping list. (b) As the goods are picked up and scanned, the corresponding information is displayed on the shopping cart's reader. (c) Since the goods in the shopping cart are pre-scanned, the transaction can be immediately completed. (d) After purchasing, the tags attached on the goods should be deactivated.

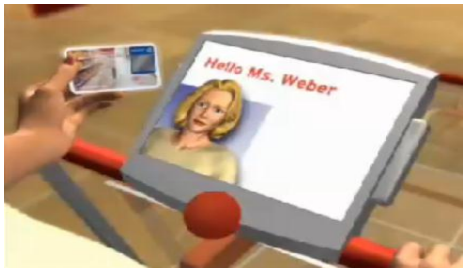
(a) Consumer logs in

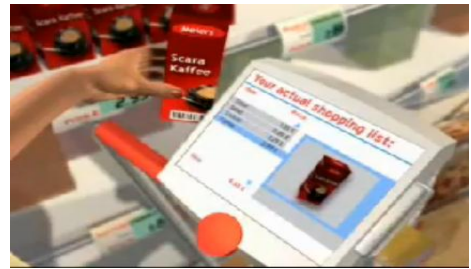
(b) Scan the goods

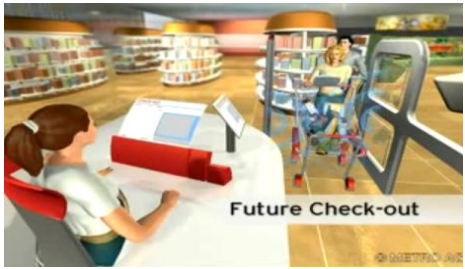
(c) Check-out

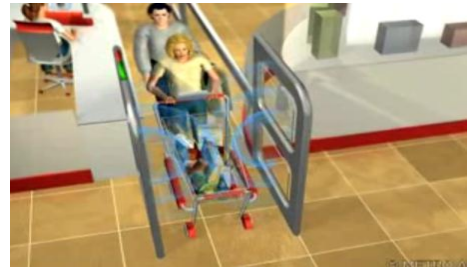
(d) Deactivate the tag

Figure 1.  the shopping scenarios in the conceptual and experimental shops. [13, 19, 20]

Now we describe how our scheme works in such environment. For making the transaction more secure and efficient, a shared secret is secure delivered to the membership card and the tags attached on the purchased goods. Based on the membership card is involved in our design, the purchased goods' tags could be deactivated under the consumer's permission. For the activation of the purchased goods' tag for the after-sales service, it is also under the owner's permission. We divide our scheme into three phases: shopping phase, de-activate in purchasing phase, and activate in after-sales service phase. There are four roles involved in our scheme as follows.

(1)Server

The server is not only the goods database but also the tag management center.

(2)RFID reader

RFID readers are set up at after-sales service center of the shopping mall. RFID readers are also set up in each intelligent shopping cart.

(3)Membership card with embedded RFID tag (m- type tag)

The membership card embedded a RFID tag is issued by the shopping mall for each member.

(4)RFID tag attached to goods (g-type tag)

The goods in the shopping mall should be embedded with RFID tags.

The notations of our proposal are listed as follows.

| | |
|---|---|
| $h()$ | One-way hash function |
| $\oplus$ | Exclusive-or |
| $pk$ | The server's public key |
| $sk$ | The server's private key |
| $E_x()$ | The encryption function with key x |
| $D_x()$ | The decryption function with key x |
| $K_{xi}$ | Secret key of the RFID tag, $K_{ci}$ means the m-type tag's secret key, $K_{gi}$ means the g-type tag's secret key. |
| $T_{xi}$ | Encrypted secret key $T_{xi} = E_{pk}(K_{xi})$, $T_{ci}$ means the m-type tag's encrypted key, $T_{gi}$ means the g-type tag's encrypted key. |
| $K_{cg}$ | The deactivate/activate key is generated by the m-type tag for deactivating/activating the g-type tag for each transaction. |
| $e_i$ | The secret seed is recorded in the m-type tag for the deactivate/activate key's generation. |
| $D_j$ | The transaction time is another factor for generating the deactivate/activate key in each transaction. |
| $w$ | The shopping token is randomly generated by the server for the authenticated m-type tag and g-type tags. It is the share secret between the m-type tag and the g-type tags in the shopping. |
| $n_i$ | The shopping pseudonym is randomly generated by the server for the authenticated user. |
| $V_g$ | The signed receipt includes serial number of the shopping mall, goods' price, transaction time and the signature of above messages. |

## 2.1 Shopping phase

For a RFID-equipped shopping cart is used by a consumer, it can be also used for authenticating the consumer's identity and inquiring the RFID-tagged goods' information. In our scheme, the m-type tag and g-type tag have the same authentication process. After the verification, all of the m-type tag and the g-type tags get a common shopping-token. Besides, the m-type tag also gets a pseudonym for this transaction and the g-type tag acquires a signed receipt, respectively.

Step 1: The reader generates a random number $\gamma_1$ and sends it to the m-type tag.

Step 2: As the tag receiving the random number $\gamma_1$, it generates the response value $v$ by the secret key $K_{xi}$:

$$v = h(K_{xi}, \gamma_1)$$

Then the tag generates a random number $\gamma_2$ and sends it to the reader with encrypted secret key $T_{xi}$ and the response value $v$.

Step 3: Upon receiving the above messages, the reader forwards the messages $(T_{xi}, v, \gamma_2)$ to the server. And the server uses its secret key $sk$ to derive and verify the tag's secret key $K'_{xi}$.

$$K'_{xi} = D_{sk}(T_{xi})$$

$$v \stackrel{?}{=} h(K'_{xi}, \gamma_1)$$

After the tag is authenticated, the server randomly generates a shopping token $w$, and both of the new key $K''_{xi}$ and the new encrypted key $T''_{xi}$ are updated.

$$K''_{xi} = h(K'_{xi})$$
$$T''_{xi} = E_{pk}(K'_{xi})$$

For the shopping token $w$, the new key of the tag $K'_{xi}$, and the new encrypted key can be secure transmitted, the following messages $(\alpha_1, \beta_1)$ are generated.

$$\alpha_1 = h(K_{xi}, \gamma_2) \oplus (w \| K''_{xi} \| T''_{xi})$$
$$\beta_1 = h(w, K''_{xi}, T''_{xi}, \gamma_2)$$

According to the tag's type, the following process of this step can be divided into two cases:

(1) m-type tag:

The server generates a shopping pseudonym $n_i$ and records it. For the pseudonym $n_i$ can be secure transmitted, the following message $(\alpha_2, \beta_2)$ is generated.

$$\alpha_2 = h(w, \gamma_2) \oplus n_i$$
$$\beta_2 = h(n_i, r_2)$$

(2) g-type tag:

The server generates a singed receipt. For the purpose of making the singed receipt $V_g$ can be secure transmitted, the following message $(\alpha_2, \beta_2)$ is generated.

$$\alpha_2 = h(w, \gamma_2) \oplus V_g$$
$$\beta_2 = h(V_g, \gamma_2)$$

Afterward, the message $(\alpha_1, \beta_1, \alpha_2, \beta_2)$ is sent to the tag through the reader.

Step 4: As the tag receiving above message, the shopping token $w$, the new key of the tag $K''_{xi}$, and the new encrypted key $T''_{xi}$ can be derived and verified as follows.

$$(w' \| K'''_{xi} \| T'''_{xi}) = \alpha_1 \oplus h(K_{xi}, \gamma_2)$$
$$\beta_1 \stackrel{?}{=} h(w', K''_{xi}, T''_{xi}, \gamma_2)$$

Then the temporary shopping token $w'$, the new secret key $K'''_{xi}$ and the new encrypted secret key $T'''_{xi}$ are recorded( or updated) in the tag.

According to the tag's type, the following process of this step can be divided into two cases:

(1) m-type tag:

The shopping token $n_i'$ can be derived and verified as follows.

$$n_i' = \alpha_2 \oplus h(w', \gamma_2)$$
$$\beta_2 \stackrel{?}{=} h(n_i', \gamma_2)$$

Then the tag records the temporary shopping pseudonym $n_i'$.

(2) g-type tag:

The shopping token $V_g'$ can be derived and verified as follows.

$$V_g' = \alpha_2 \oplus h(w', \gamma_2)$$
$$\beta_2 \stackrel{?}{=} h(V_g, \gamma_2)$$

Afterward, the tag records the signed receipt $V_g'$.

## 2.2 Purchasing phase

After purchasing, for protecting the location privacy of the consumer, all of the g-type tags should be deactivated. The deactivate/activate key is determined by the m-type tag and secure transmitted to the g-type tag. In particular, the deactivate/activate key is not necessary to be recorded in the m-type tag in our design. It will be practical to be implemented on the kinds of memory limited RFID tag.

Step 1: The reader sends a query message to the m-type tag.

Step 2: The m-type tag generates a random number $\gamma_1$ and sends to the reader with the shopping pseudonym $n_i$.

Step 3: Upon receiving above messages, the reader forwards them to the server. According to the shopping pseudonym $n_i$, the server confirms this transaction is processed for which consumer. The corresponding secret key $K_{ci}$ is retrieved from the database. The transaction time $D_j$ will be secure transmitted to the m-type tag by the following message $(\alpha_1, \beta_1)$.

$$\alpha_1 = h(K_{ci}, \gamma_1) \oplus D_j$$
$$\beta_1 = h(D_j, \gamma_1)$$

The server also generates a random number $\gamma_2$. The messages $(\alpha_1, \beta_1)$ and the random number $\gamma_2$ are forwarded to the m-type tag through the reader.

Step 4: As the m-type tag receiving above messages, the transaction time $D_j'$ can be derived and verified as follows.

$$D_j' = \alpha_1 \oplus h(K_{ci}, \gamma_1)$$
$$\beta_1 \stackrel{?}{=} h(D_j', \gamma_1)$$

According to the transaction time $D_j'$, the m-type tag uses its secret seed $e_i$ to generate the deactivate/activate key $K_{cg}$.
$$K_{cg} = h(e_i, D_j')$$

Now, the following message $(\delta, v)$ is prepared for transmitting the deactivate/activate key $K_{cg}$ and the transaction time $D_j'$ to the g-type tags secretly, the m-type tag uses the shopping token $w$ to generate them as follows.
$$\delta = w \oplus (K_{cg} || D_j')$$
$$v = h(K_{cg}, D_j', \gamma_2)$$

For the integrity of the messages $(\delta, v)$ can be verified by the reader, the following messages $(\alpha_2, \beta_2)$ are generated.
$$\alpha_2 = h(K_{ci}, \gamma_2) \oplus (\delta || v)$$
$$\beta_2 = h(\delta, v, \gamma_2)$$

Then the m-type tag emits the messages $(\alpha_2, \beta_2)$ to the reader.

Step 5: Upon receiving above messages, the reader forwards them to the server. Then the messages $(\delta' || v')$ can be derived and verified as follows.
$$(\delta' || v') = \alpha_2 \oplus h(K_{ci}, \gamma_2)$$
$$\beta_2 \stackrel{?}{=} h(\delta', v', \gamma_2)$$

Then the messages $(\delta', v')$ and the random number $\gamma_2$ should be broadcasted to the g-type tags in the shopping cart.

Step 6: As a g-type tag receiving the messages, the deactivate/activate key $K_{cg}'$ and the transaction time $D_j''$ can be derived and verified as follows.
$$(K_{cg}' || D_j'') = \delta' \oplus w$$
$$v' \stackrel{?}{=} h(K_{cg}', D_j'', \gamma_2)$$

After the deactivate/activate key $K_{cg}'$ and the transaction time $D_j''$ are recorded, the g-type tag deactivates itself.

**2.3 After-sales service phase**

For getting after-sales service, the g-type tag should be activated and verified. The procedure is taken by the cooperation of the corresponding m-type tag which has the ability to generate the deactivate/activate key.

Step1. The reader sends a query message to the g-type tag.

Step2. Under deactivated status, the g-type tag just response the non-characteristic value $D_j$ and a random number $\gamma_1$ to the reader.

Step3. Upon receiving above messages, the reader forwards them to the m-type tag.

Step4: According to the value $D_j$, the m-type tag uses its secret seed $e_i$ to generate the deactivate/activate key $K_{cg}$.

$$K_{cg} = h(e_i, D_j)$$

According to the challenging message $\gamma_1$, the response $\delta$ for the g-type tag can be generated as follows.

$$\delta = h(K_{cg}, \gamma_1)$$

For the response $\delta$ can be secure transmitted, the following message $(\alpha_1, \beta_1)$ is generated.

$$\alpha_1 = h(K_{ci}, \gamma_1) \oplus \delta$$
$$\beta_1 = h(\delta, \gamma_1)$$

Another response value $v_1$ is necessary to be generated for authentication.

$$v_1 = h(K_{ci}, \gamma_1)$$

Afterward, the m-type tag generates a random number $\gamma_2$ and sends it to the reader with the encrypted key $T_{ci}$, the response $v_1$, and the messages $(\alpha_1, \beta_1)$.

Step 5: Upon receiving above messages, the reader forwards them to the server. And the server uses its secret key $sk$ to derive and verify the m-type's secret key $K'_{ci}$.

$$K'_{ci} = D_{sk}(T_{ci})$$
$$v_1 \stackrel{?}{=} h(K'_{ci}, \gamma_1)$$

After the m-type tag is authenticated, both of the new key $K'_{xi}$ and the new encrypted key $T'_{xi}$ are updated.

$$K'_{ci} = h(K_{ci})$$
$$T'_{ci} = E_{pk}(K'_{ci})$$

Now, the server will derive and verify the response $\delta'$ for the g-type tag as follows.

$$\delta' = h(K_{ci}, \gamma_1) \oplus \alpha_1$$
$$\beta_1 \stackrel{?}{=} h(\delta', \gamma_1)$$

Then the response $\delta'$ is forwarded to the g-type tag by the reader.

Step 6: As the g-type tag receiving the response message, it can be verified by the recorded deactivate/activate key $K_{cg}$ as follows.

$$\delta' \stackrel{?}{=} h(K_{cg}, \gamma_1)$$

Based on the response message is confirmed with embedded the corresponding deactivate/activate key, the g-type tag activates itself.

Now, the g-type tag will generate the response $v_2$ for authentication.

$$v_2 = h(K_{gi}, r_1)$$

The g-type tag emits the encrypted key $T_{gi}$, the response value $v_2$ and the signed receipt $V_g$ to the reader.

Step 7: Upon receiving above messages, the reader forwards the messages to the server. And the server uses its secret key $sk$ to derive and verify the g-type tag's secret key $K'_{gi}$.

$$K'_{gi} = D_{sk}(T_{gi})$$
$$v_2 \stackrel{?}{=} h(K'_{gi}, \gamma_1)$$

After the g-type tag is authenticated, the after-sales service should be provided as it is within the warranty period that can be proved by the signed receipt $V_g$. For finalizing this session, the new secret key $K'_{ci}$ and the new encrypted key $T'_{ci}$ which generated in step 5 should be also updated in the m-type tag. The following acknowledgment messages $(\alpha_2, \beta_2)$ are generated.

$$\alpha_2 = h(K_{ci}, \gamma_2) \oplus (K'_{ci} || T'_{ci})$$
$$\beta_2 = h(K'_{ci}, T'_{ci}, \gamma_2)$$

Then the messages $(\alpha_2, \beta_2)$ are forwarded to the m-type tag by the reader.

Step 8: As the m-type tag receiving above message, the new secret key $K'_{ci}$, and the new encrypted key $T'_{ci}$ can be derived and verified as follows.

$$(K''_{ci} || T''_{ci}) = \alpha_2 \oplus h(K_{ci}, \gamma_2)$$
$$\beta_2 \stackrel{?}{=} h(K''_{ci}, T''_{ci}, \gamma_2)$$

Finally, the new secret key $K''_{ci}$ and the new encrypted secret key $T''_{ci}$ are also updated in the m-type tag.

## 3. Analysis

### 3.1 Criteria analysis of after-sale service

For supporting after-sales service in a RFID system, here are a few suggestive such applications [1, 7, 8, 10]:

- Stores may wish products to have tags scannable if the products are returned as defective [10].
- RFID tags could carry relevant information to achieve the goal of "receipt-less" item returns [7, 8].
- A merchant may wish to scan consumers for marketing purposes [10].
- Products may need to scannable so they may be categorized for recycling purposes [10].
- The reactivation would require the permit of the owner of the RFID tag [1, 8].

#### 3.1.1 Stores may wish products to have tags scannable if the products are returned as defective.

In our scheme, the g-type tags attached to the goods are deactivated after purchasing. The g-type tags can be only activated by the corresponding m-type tag as the goods are returned to the shop for asking after-sales service. As the m-type tag gets the transaction time value $D_j$ from the g-type tag, the correct deactivate/activate key $K_{cg} = h(e_i, D_j)$ can be generating by the corresponding seed $e_i$.

$$\delta = h(K_{cg}, \gamma_1)$$
$$\alpha_1 = h(K_{ci}, \gamma_1) \oplus \delta$$
$$\beta_1 = h(\delta, \gamma_1)$$

The server will derive and verify the response $\delta$ by the m-type tag's key $K_{ci}$ as follows.

$$\delta = h(K_{ci}, \gamma_1) \oplus \alpha_1$$
$$\beta_1 \stackrel{?}{=} h(\delta, \gamma_1)$$

Then the response $\delta$ is forwarded to the g-type tag by the reader, and be verified by the recorded deactivate/activate key $K_{cg}$ as follows.

$$\delta \stackrel{?}{=} h(K_{cg}, \gamma_1)$$

Based on the response message is confirmed to be embedded with the corresponding deactivate/activate key which is protected by the one-way hash function, the g-type tag activates itself. Then the g-type tag can be scannable again.

### 3.1.2 RFID tags could carry relevant information to achieve the goal of "receipt-less" item returns.

For the purchased goods, the signed receipt $V_g$ (which including the price, the shop's number, the transaction time, the discount, and the signature) is recorded in each g-type tag. As the g-type tag is activated again, the signature in the receipt can be verified as a "receipt-less" proof.

### 3.1.3 **A merchant may wish to scan consumers for marketing purposes**.

In our design, the m-type tag can be only scanned by the membership card issuer, such design will meet the requirement for merchants' wish to scan consumers, but consumers' privacy is guaranteed. As a m-type tag is queried by a challenge value $\gamma_1$, it just response its encrypted key and a hash value $v$ which is generated as follows.

$$v = h(K_{ci}, \gamma_1)$$

The response messages $T_{ci}$ and $v$ can be only useful for the membership card issuer. Using the correct private key $sk$, the issuer can authenticate the tag as follows.

$$K'_{ci} = D_{sk}(T_{ci})$$
$$v \stackrel{?}{=} h(K'_{ci}, \gamma_1)$$

Clearly, it is meaningless as the m-type tag scanned by others to get the encrypted value $T_{ci}$ and the hash value $v$.

Moreover, our scheme considers the equation from the consumer's point of view, it provides value to consumers: peace of mind, convenience, improved service. Therefore, to deploy such RFID mechanism should be helpful for marketing.

### 3.1.4 Products may need to scannable so they may be categorized for recycling purposes.

In our design, the g-type tags can be also used for the recycle management just after activating again.

### 3.1.5 The reactivation would require the permit of the owner of the RFID tag.

According the researches [1, 8] have discussed that the law should keep the shops from supervising, tracking or infringing the consumers' privacy. In other words, the reaction of the g-type tag should be only the owner desire. For ensuring g-type tags is activated under the owner's permission, the RFID membership card is involved in the after-sales service phase as we mentioned in section 3.1.1. Of course, the service staff of the shop has the responsibility to confirm the membership card is used by the owner.

**3.2 Security analysis**

In our scheme, the most concerned issue of location privacy is guaranteed. Besides, the general security issues such as confidentiality, data integrity, mutual authentication, replay attack, and non-repudiation are analysed in this section.

**3.2.1  Location privacy:**

In our design, the g-type tag is immediately deactivated after purchasing. For any query, the deactivated g-type tag just responses a non-characteristic transaction time $D_j$. The value $D_j$ is not useful for tracking the tag. It is not unique for each tag, since there maybe a lot of transactions happen at the same time. Therefore, those g-type tags carried by people will not cause the location privacy problem.

However, may the m-type tag carried by people cause the tracking problem? The most facile solution to avoid disclosing the consumer's location is that the consumer put the m-type tag in a Faraday Cage wallet to block any query. However, as the m-type tag has to be taken out for authentication, the m-type tag's response has to be avoided being associated with the individual consumer. Based on the pseudonym approach is applied in the authentication, the encrypted pseudonym $T_{ci}$ is not helpful for the attacker to understand the linkage of the m-type tag. Only the server can decrypt $T_{ci}$ to authenticate the m-type tag. After the authentication, the encrypted pseudonym is updated to be $T_{ci}' = E_{pk}(K_{ci}')$ which is securely transmitted to the m-type tag for symmetric update. After the tag updating the new encrypted key $T_{ci}''$, the response will be different in next authentication. Therefore, the m-type tag will not be tracked or monitored by the attacker on the authentication process.

**3.2.2  Confidentiality:**

In our scheme, the confidential messages are always under the protection of the one-way hash function and exclusive-or operations for transmitting. Only the target entity uses the corresponding secret key to derive the message. Therefore, the confidentiality is guaranteed in our design.

### 3.2.3 Data integrity:

For preventing possible malicious altering or transmission errors, the message integrity verification are always involved in each transmitting. Therefore, the integrity of the transmitted messages is guaranteed in our design.

### 3.2.4 Mutual authentication:

Mutual authentication refers to both entities in a communications link authenticate each other suitably. In RFID system, it mentions a tag authenticating itself to a server and vice-versa, therefore the both parties' identity are verified. In our scheme, the tag and the server can use the common secret to authenticate each other based on the design of twice challenge-response. For instance, in the shopping phase, the mutual authentication between the tag and the server is achieved according to the steps described in Figure 2.

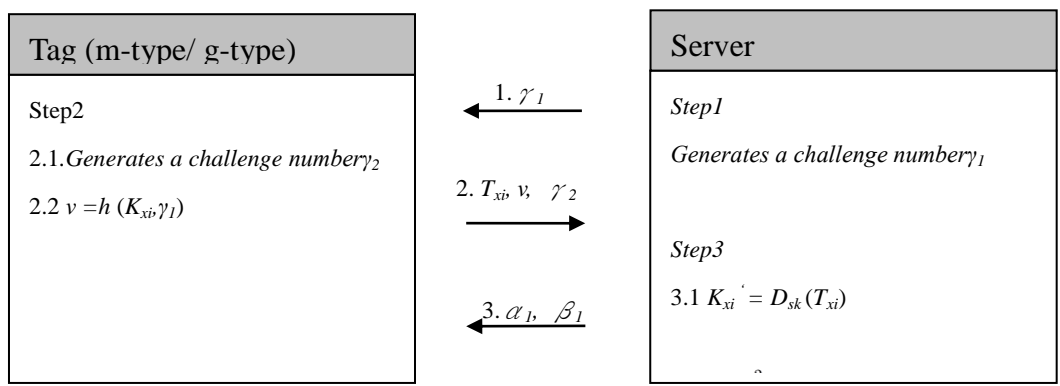

Figure 2.   The authentication scenarios in the shopping phase

Upon receiving the challenge value $\gamma_1$ from the server, the tag generates a response value $v$ embedded with the receiving challenge and its secret key $K_{xi}$ which is a secret shared with the server. The tag's challenge $\gamma_2$ is also generated immediately. As the tag's secret key $K_{xi}'$ is decrypted by the server, the tag's response can be verified by the tag's challenge value $\gamma_1$ as follows. $v \stackrel{?}{=} h(K_{xi}', \gamma_1)$ After, the server's response ($\alpha_1$, $\beta_1$) which embedded the common secret key $K_{xi}$ and the tag's challenge value $\gamma_2$ are generated. As the updated values ($w'//K_{xi}''//T_{xi}''$) is derived by the tag, the server's response can be verified by the challenge value $\gamma_2$ as

follows. $\beta_1 \stackrel{?}{=} h(w', K_{xi}'', T_{xi}'', \gamma_2)$

Similarly, the mutual authentication between the m-type tag and the server is also achieved in the after-sales service phase. Only after the authentication, the m-type tag can be used by the owner to activate the g-type tags attached on the purchased goods. Then the activated g-type tags can be authenticated by the server for providing after-sales service.

### 3.2.5 Resist replay attack:

As a tag authenticating itself to the server, its response is generated according to the random challenge determined by the server. For instance, in the step 2 of the shopping phase, a tag generates a response value $v$ which is included the reader's challenge value $\gamma_1$. Clearly, the response cannot be reused by anyone for cheating the server. Similarly, the server has to authenticate itself to the tag, the replay attack can be resisted due to the authentication is also based on the design of challenge-response.

### 3.2.6 Non-repudiation:

In our scheme, the signed receipt is recorded in each g-type tag attached on the purchased goods. The price, the shop's identity, the purchase date, even the discounts are included in the receipt. With the corresponding signature is also recorded in the g-type tag, it cannot be denying for the goods sold by the shop.

### 4. Conclusion

The development of RFID-based systems is an overall trend of the world, it not only provides the commercial benefits of the company, but also could satisfy the consumers' requirement. For increasing the consumers' acceptance as applying RFID in the retailers, Bruce Eckfeldt[4] proposed three key points: peace of mind, consumer convenience, and improved service. For making consumers feel relieved in using our scheme, the tracking can be prevented by deactivating the tag of purchased goods. Consumers can enjoy the convenient service without additional burden such as extra devices or complicated processes in our scheme. In fact, the confirm process for after-sales service is improved by using more efficient "receipt-less" proof. We believe our design is a potential and useful scheme for improving the customer-service and consumer shopping experience.